\documentclass[conference, a4paper]{IEEEtran}
\ifCLASSINFOpdf
\else
\fi
\usepackage{graphicx}
\usepackage{todonotes}
\usepackage[utf8]{inputenc}
\usepackage{amsmath,amssymb,amsfonts}
\usepackage{textgreek}
\usepackage{graphicx}
\usepackage{color,bm}
\usepackage{wasysym}
\usepackage{tikz}
\usepackage{cite}
 \usepackage{array,multirow}
 \usepackage{float}
 \usepackage{subfig}
 \usepackage{balance}
\DeclareGraphicsExtensions{.tif}
\usepackage{amsmath}
\usepackage{geometry}
\newgeometry{left=1.91cm,right=1.31cm,top=4.1cm,bottom=1.91cm}

\begin{document}
%
\title{\LARGE \textbf{The State-of-the-Art of Coordinated Ramp Control with Mixed Traffic Conditions}}

\author{Zhouqiao~Zhao,
        Ziran~Wang,
        Guoyuan~Wu,
        Fei~Ye,
        and~Matthew~J.~Barth
        
\IEEEauthorblockA{\\Center for Environmental
Research and Technology\\
University of California, Riverside, CA, USA 92521\\
email: zzhao084@ucr.edu; zwang050@ucr.edu; gywu@cert.ucr.edu; fye001@ucr.edu; barth@ece.ucr.edu}}



%


\maketitle

\begin{abstract}
Ramp metering, a traditional traffic control strategy for conventional vehicles, has been widely deployed around the world since the 1960s. On the other hand, the last decade has witnessed significant advances in connected and automated vehicle (CAV) technology and its great potential for improving safety, mobility and environmental sustainability. Therefore, a large amount of research has been conducted on cooperative ramp merging for CAVs only. However, it is expected that the phase of mixed traffic, namely the coexistence of both human-driven vehicles and CAVs, would last for a long time. Since there is little research on the system-wide ramp control with mixed traffic conditions, the paper aims to close this gap by proposing an innovative system architecture and reviewing the state-of-the-art studies on the key components of the proposed system. These components include traffic state estimation, ramp metering, driving behavior modeling, and coordination of CAVs. All reviewed literature plot an extensive landscape for the proposed system-wide coordinated ramp control with mixed traffic conditions.
\end{abstract}


%
\IEEEpeerreviewmaketitle

\section{Introduction}
With ever increase of travel demands, freeways play a significant role in facilitating the transportation-related activities. As an indispensable part of the freeway system, on-/off-ramps and the associated controls have attracted much attention from worldwide researchers for the following reasons: 
\begin{itemize}
    \item The ramp merging areas have consistently witnessed potential conflicts between on-ramp vehicles and main-line vehicles; 
    \item Uncontrolled inflow traffic to the freeway network can lead to congestion; 
    \item Due to limited vision range and uncoordinated merging behaviors with other vehicles, the drivers may experience unnecessary acceleration/deceleration, thus resulting in excessive energy consumption and pollutant emissions. 
\end{itemize}
Therefore, ramp metering becomes a key element of advanced demand and traffic management strategy for freeways \cite{fhwa2014rampmetering}, whose major goal is to mitigate mainline congestion by controlling the entrance rate from on-ramps. Traditionally, ramp metering is designed for conventional vehicles with human drivers. It utilizes the traffic signals (green and red phases) installed at on-ramps to regulate the inflow rate to the freeway network. For each green phase, one or more vehicles are allowed to enter the mainline, and the metering rate depends on the estimated traffic states in real time, such as lane occupancy, average speed, and length of on-ramp queue. This information is usually collected with fixed-location sensors, \textit{e.g.,} inductive loop detectors and cameras. 

Connected and automated vehicle (CAV) technology has been studied extensively in the last decade. The idea of automated vehicle based freeway system can be traced back to 1970\cite{Fenton1970}. With the on-board sensing system, CAVs can perceive the surrounding environment. With the communication abilities between CAVs and roadside infrastructures and among CAVs, detailed and accurate traffic information can be shared. Thus, higher resolution of system states can be estimated for improving the adaptability and system-wide optimality of ramp metering strategies. Furthermore, the controllability of the automated vehicles enables the centralized or distributed coordination on the maneuvers of a number of vehicles for better merging. This may unlock abundant opportunities for more efficient and more delicate ramp control protocols on an individual vehicle basis. Nevertheless, assuming all vehicles are CAVs may be too ambitious for the near future, and it is expected that CAVs would have to coexist with other conventional vehicles for another decade(s). Therefore, the solution to a mixed traffic would be of much realistic value although it still remains a major challenge.

The goal of this paper is twofold: a) to propose an innovative system architecture for cooperative ramp control in the mixed traffic environment; and b) to review relevant studies on the key components of the proposed system. The remainder of this paper is organized as follows. In Section \ref{SA}, we propose a hierarchical architecture for the mixed traffic coordinated ramp control system. In Section \ref{TSE}, we survey the area of traffic state estimation. In Section \ref{RM}, ramp metering algorithms for conventional vehicles are reviewed. A survey of driving behavior modeling is presented in Section \ref{DBM}. Section \ref{CAV} reviews the coordination algorithms for CAVs, especially at freeway ramp areas. Section \ref{Conclusion} concludes the findings from literature review and discusses some future research directions.

\section{System architecture}\label{SA}
\begin{figure}
    \centering
    \includegraphics[width=1.0\columnwidth]{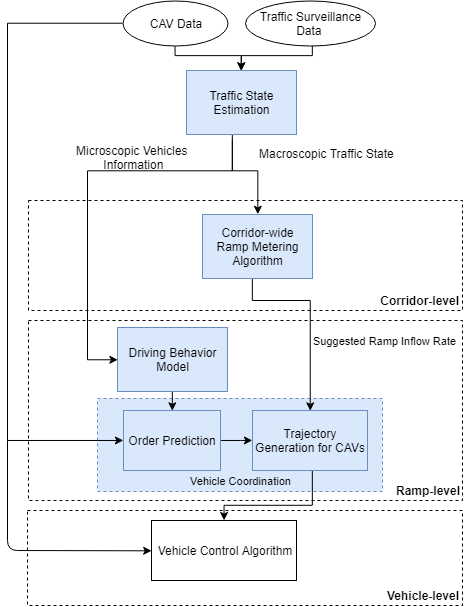}
    \caption{Demonstration of the System Architecture}
    \label{fig:flow}
\end{figure}

The typical freeway system contains multiple on-ramps and off-ramps. By leveraging CAV technology, we can better estimate and predict the traffic states in real time, and develop a ramp control system that can: 1) improve the operational efficiency of entire freeway by cooperatively regulating the\newgeometry{left=1.91cm,right=1.31cm,top=3.6cm,bottom=1.91cm}\noindent inflow rates of all ramps; 2) mitigate safety concerns at the ramp merging areas by coordinating the movements of vehicles on both mainline and ramp; and 3) reduce the excessive vehicular energy consumption and emissions by smoothing the longitudinal maneuvers of CAVs and other traffic. Because of the involvement of conventional human-driven vehicles, design of the trajectories for CAVs should predict the behavior of conventional vehicles based on certain models such as car-following and merge gap acceptance. Fig. \ref{fig:flow} illustrates the system architecture of the proposed cooperative ramp control system for mixed traffic. As shown in the figure, the hierarchical system can be divided into a real-time data processor and a 3-level structure.

The first key component of the system is \emph{traffic state estimation}. Because of the mixed traffic condition, not every vehicle can be observed exhaustively. Therefore, the input data fed into the module includes CAV data (such as location and instantaneous speed), and traffic surveillance data (such as vehicle counts and lane occupancy). This data partially reflect the real-time traffic conditions. The goal of traffic state estimation is to provide accurate prediction of the current traffic information both microscopically and macroscopically based on apriori knowledge and partial observation\cite{Seo2017}. The Section \ref{TSE} provides a comprehensive review of current research of traffic state estimation.

At the corridor-level, the ramp metering algorithm calculates system-wide optimal inflow rate for each on-ramp, given the estimate of macroscopic traffic states. The resultant inflow rate serves as the constraint for boundary control at lower level (\textit{i.e.}, ramp-level). 
For mixed traffic, the conventional ramp metering algorithms need to be further extended to consider CAVs in the context, where a hybrid ramp control strategy should be developed. In Section \ref{RM}, we will review not only corridor-wide ramp metering algorithms but also some well-known localized algorithms such as ALINEA\cite{Papageorgiou1991}.

The ramp-level module coordinates the maneuvers of CAVs and human-driven vehicles at the merging area and regulates the ramp inflow rate based on the output from the corridor-level control. Desired CAVs trajectories are calculated by a centralized controller. To complete this task, the controller should keep tracking the human-driven vehicles and predicting their behaviors, and the trajectories of CAVs should be adjusted accordingly. Therefore, two key components will be reviewed and presented in Section \ref{DBM} and \ref{CAV}. Section \ref{DBM} summarizes the research on driving behavior modeling for conventional vehicles, while Section \ref{CAV} presents the recent findings of cooperative merging for CAVs. Omidvar \textit{et al.}\cite{Omidvar2018} observed that the human-driven tend to stay a constant speed, which may simplify the modeling of conventional vehicles' dynamics. For the merging protocol of mixed traffic, we propose a two-step centralized control scheme (for CAVs) as shown in figure \ref{fig:flow}. The first step is to predict the order of the vehicles based on the estimated time-to-arrival (TOA) at the merging point. Some existing algorithms use heuristic rules like first-come-first-serve\cite{Rios2017_ramp}. Then, a centralized controller can be designed to govern the movement of CAVs when interacting with conventional vehicles, without compromising the safety.

Once the trajectory is calculated for the involved CAV, a lower-level controller (at the vehicle dynamics level) is designed for trajectory tracking. Please note that the power-train control of the CAVs is not up for discussion in this paper. 

\section{Traffic State Estimation}\label{TSE}
Accurate measurement and estimation of prevailing traffic conditions are the foundation of effective ramp control system. Since it is difficult and expensive to obtain complete information on the traffic (\textit{e.g.}, 100\% penetration rate of connected vehicles), estimation of traffic states, such as flow, density, and speed, from partially observed traffic data plays an important role. Seo \textit{et al.} performed a comprehensive survey about traffic state estimation which provides a guideline into this field \cite{Seo2017}. The categories listed below are on the basis of their suggestion.

At the macroscopic scale, road networks are divided into several segments without further merging or diverging, which are called links. Inside each link, the traffic states can be considered to be homogeneous. Usually, \emph{flow} (veh/hr), \emph{density} (veh/km), and \emph{speed} (km/hr) are three key variables used for traffic flow management. There might be other equivalent variables. For example, some ramp metering algorithms use occupancy as system input, which is quite comparable to \emph{density} with the assumption or estimation of vehicle length.

Raw data available for estimation can be categorized as: \emph{stationary} and \emph{mobile}. The stationary data is collected by fixed location sensors such as inductive loop detectors, micro-wave radars on roadside, and surveillance cameras. The mobile data is collected by moving vehicles equipped with sensors such as GPS, on-board radar, and camera. The emergence of CAVs on roads can provide a significant amount of mobile sensor data.

Traffic flow model is widely used for traffic state estimation. The models are usually based on physical and empirical relations. Borrowing the idea of hydrodynamic theory, fundamental diagram depicts the relationship between density and speed, and/or density and flow\cite{greenshields1935, NEWELL1993281}. Another dynamic representation model with the hydrodynamic theory is cell transmission model (CTM)\cite{DAGANZO1994}, which utilizes the discrete analog of the differential equations arising from a special case of the hydrodynamic model of traffic flow. Because of the conservation law for the traffic hydrodynamic, the aggregated behavior of traffic is depicted by partial differential equations. Lighthill-Whitham-Richard (LWR) model is the most commonly used first-order model\cite{Lighthill1955, Richards1956}, and there are many higher-order models such as Payne-Whitham (PW) model\cite{Payne1971, Whitham1974}.

\subsection{Model-based Approaches}
Model-based approaches are widely used in the field of traffic state estimation, where aforementioned traffic flow models are applied. The parameters of the models are usually calibrated via historical data from the field. After the calibration, target data is fed into the model to estimate the traffic states. Coifman proposed a method of estimating microscopic vehicle information (\textit{i.e.}, travel times and trajectories) using LWR model and loop detector data\cite{COIFMAN2002}. Wang and Papageorgiou estimated macroscopic traffic via Extended Kalman Filter (EKF)\cite{WANG2005}. However, these approaches may fail if inappropriate models are adopted.
Furthermore, as the parameters are calibrated by specific data sets, the model-based approaches may not be so adaptive to the drastic changes of traffic conditions.

\subsection{Learning-based Approaches}
Different from model-based approaches, no empirical traffic flow models are used in learning-based approaches. Learning-based models are trained through statistical methodologies or machine learning techniques given a large amount of historical data. A typical learning-based estimation approach was proposed by Tak \textit{et al.}\cite{Tak2016}, in which k-nearest neighbors (KNN) algorithm was applied. Because of the challenge to well interpret the features of learning scheme, it is difficult to analytically comprehend how the entire system works even though the results may be very attractive. Additionally, similar to the model-based approaches, the dependency of trained data set makes the system not flexible enough to transfer to other untrained scenarios.

\subsection{Streaming-data-driven Approaches}
Without using neither empirical models nor historical data, streaming-date-driven approaches only rely on real-time data and "weak" assumptions. Therefore, it is more robust to different traffic conditions. The "weak" assumption such as \emph{Conservation Law} is generally reasonable, considering the physical constraints. A recent research was performed by Florin \textit{et al.}\cite{Florin2017}, who presented mobile observer method with aggregated information of number of overtaking maneuver of vehicles. The increasing number of CAVs and connected vehicles on freeways can provide a large amount of streaming data for traffic state estimation, which makes it possible to consider mixed traffic scenarios. Bekiaris-Liberis \textit{el al.} proposed a mixed traffic state estimation method with streaming-data-driven approach utilizing only average speed measurements reported by connected vehicles and a minimum number (sufficient to guarantee observability) of spot-sensor-based total flow measurements \cite{Bekiaris-Liberis2016}.

\section{Ramp Metering}\label{RM}
A ramp meter is a traffic signal on a freeway on-ramp area that is used to regulate the inflow rate of vehicles onto the freeway. Although there are some review papers about ramp metering algorithms in the past few decades, such as \cite{Shaaban2016, Berkeley2001, Bogenberger1999}, latest progress in this area especially related to machine learning technique was not included. In this section, we will divide the up-to-date ramp metering algorithms into three categories: rule-based, control-based, learning-based, and review them respectively. 

\subsection{Rule-based Approaches}
A typical rule-based approach contains a hierarchy logic to adjust the ramp rate internally and coordinately. In general, the rule-based algorithms are easy to define, modify, and are relatively computational efficient. They simplify the nonlinear feature of the traffic, which is complex to be modeled accurately without hard assumptions. 

A widely used rule-based method to achieve system-wide coordination is competitive approach, which usually contains two competitive controllers running in parallel, where the more restrictive metering rate would be chosen. Additional adjustment could be introduced to address other constrains like queue effects. Bottleneck Algorithm \cite{Jacobson1989} used upstream occupancy and bottleneck data as system inputs, and the algorithm contains coordinated bottleneck controllers and local controllers. 
 Similar to the Bottleneck Algorithm, two-module structure were proposed in the System-Wide Adaptive Ramp Metering (SWARM) Algorithm \cite{Paesani1997} as well. SWARM made decision based on estimated lane density using Kalman filter and linear regression. The local controller (called SWARM2) calculated metering rates by preserving headway which was converted from estimated local density, and the coordinated controller (called SWARM1) calculated metering rates to adjust the current density to the desired value.

ZONE\cite{Stephanedes1994} Algorithm was another well known rule-based approach which segmented freeway into zones, and it used \emph{Conservation Law} to model the volume change of the traffic. By integrating the real-time measurement on upstream mainline volume with historical data, the algorithm chose a predefined metering rate accordingly. Fuzzy Logic\cite{Taylor1998, Taylor2000} algorithm was developed in 1998, which converted empirical knowledge into finite fuzzy rules. The proper choice of rules might lead to a robust system, but it could be overwhelming to identify appropriate rules for a system-wide ramp metering. Advanced Real-time Metering System (ARMS)\cite{Liu1994a} was made up of free-flow control, congestion prediction, and congestion reduction. The algorithm maximized the throughput and tried to reduce the risk of congestion, and Origin-Destination (O-D) information was used to distribute the calculated ramp volume to each ramp. Dynamic Ramp Metering\cite{Chen2017} is a hierarchical coordinated control of ramps that contains state estimation, O-D prediction, local control and area-wide control to minimize the total system travel time.
\subsection{Control-based Approaches}
Control-based approaches introduce automatic control techniques such as feedback control into the ramp metering system. Unlike the rule-based approaches whose strategies may be sophisticated, control-based ones are usually succinct, robust and efficient. However, the nonlinearity of the traffic dynamics pose some challenges for these approaches. 

ALINEA was first proposed by Papageorgiou \textit{et al.}\cite{Papageorgiou1991} in 1991. By introducing the feedback into the control design, the algorithm is robust to the disturbance. Papageorgiou \textit{et al.} also provided a guideline for the design of ALINEA ramp metering system\cite{Papageorgiou2008}. The algorithm uses downstream occupancy as system inputs, controls the metering rates in response to the change of occupancy, and regulates them to the desired level. Although ALINEA is a local ramp metering algorithm, it draws much attention due to its simplicity, stability, and efficiency. Therefore, a large number of variations of ALINEA have been proposed to adapt to more complicated scenarios (e.g., METALINE, FL-ALINEA, UP-ALINEA, X-ALINEA/Q, PI-ALINEA)\cite{Papageorgiou1990, Berkeley2001, Smaragdis2004, Jiang2012, Kan2016} or to extend for achieving system-wide benefits (HERO)\cite{Papamichail2010}.

Linear-quadratic (LQ) feedback control algorithms are another type of control-based methods proposed by Isaksen and Payne\cite{Isaksen1973} and Golstein and Kumar\cite{Goldstein1982}, which optimize the system performance such as throughput (vehicle miles per hour). In order to minimized total delay of the freeway network, Model Predictive control (MPC) approaches were applied by some researchers. Hegyi \textit{et al.}\cite{Hegyi2005} proposed a MPC scheme in a rolling horizon framework to coordinate the variable speed limits and ramp metering. To coordinate the traffic flows in an urban network with both freeways and arterial, Haddad \textit{et al.}\cite{Haddad2013} developed an MPC-based algorithm for perimeter control including ramp metering strategies. Stratified Ramp Metering Algorithm was proposed by Geroliminis\textit{et al.}\cite{Geroliminis2011}, where a more accurate density estimation algorithm was developed. The metering rates were calculated to delay the onset of the breakdown and to accelerate system recovery. The algorithm could be regarded as a extended version of Stratified Zone Algorithm (SZM)\cite{Geroliminis2013} which aims to maximize the network throughput. The MPC scheme was also used to solve the optimization problem.

\subsection{Learning-based Approaches}
With the rapid advance in machine learning techniques such as artificial neural network (ANN) and reinforcement learning, more and more learning-based approaches for ramp metering emerge recently. Because of the nonlinear feature of traffic dynamics, learning-based approaches seem to be a shortcut to achieve better performance.

Zhang \textit{et al.} proposed a local freeway ramp metering using ANN\cite{Zhang1997}. The controllers calculate the proportional-integral (PI) feedback gain using multi-layer feed-forward (MLF) neural network which can be tuned by both historical data and on-line streaming data. Coordinated Ramp metering algorithm using ANN was firstly proposed by Wei and Wu\cite{Wei1996}. The coordinated metering rate was trained by a traffic simulation model and expert system.

The recent breakthroughs of Reinforcement Learning (RL) provide another promising direction for ramp metering. Numerous RL-based algorithms have sprung up in the past five years. A local ramp metering algorithm was developed by Lu and Huang\cite{Lu2017a}. Fares \textit{et al.}, and Lu \textit{et al.} proposed a coordinated ramp metering algorithm using Q-Learning technique\cite{Fares2015, Lu2017}. Belletti \textit{et al.} proposed an expert level ramp metering control based on multi-task deep reinforcement learning\cite{Belletti2018}, and Schmidt-Dumont \textit{et al.} combined ramp metering with variable speed limits based on decentralised reinforcement learning\cite{Schmidt-Dumont2015}.

\section{Driving Behavior Modeling}\label{DBM}
Modeling and predicting the driving behavior of conventional human-driven vehicles are essential for designing the motion behavior of CAVs in mixed traffic conditions. As the foundation of microscopic traffic models, car-following (CF) logic describes the longitudinal interactions between vehicles assuming there is no lane changing or overtaking. Over the past decades, a considerable number of car-following models have been proposed and developed \cite{CF-review, RecentReview2014}. For instance, Gipps model \cite{GIPPS}, Krauss model \cite{Krauss} and intelligent driver model (IDM) \cite{IDM} were well-developed to address the speed adjustment according to the principle of collision avoidance between vehicles. A comprehensive comparative study of car-following models used in the state-of-the-art microscopic traffic simulators was conducted in \cite{CF1}. More recently, to improve the traffic flow stability, an anticipation optimal velocity model (AOVM) was proposed by Peng \textit{et al.} considering the anticipation effect of optimal velocity\cite{NewCF2013}. Given that human factors plays an essential role in driving behaviors especially under complex traffic conditions, notable efforts have been made to integrate human factors into the conventional CF model in order to describe more realistic driving behavior \cite{RecentReview2014}. For example, an adaptive neural-fuzzy inference system (ANFIS) was proposed in \cite{ANFIS2011} that integrated human expert knowledge and neural network to adapt the vehicle speed for car-following and collision prevention. Instead of assuming constant reaction time, Khodayari \textit{et al.} proposed a artificial neural network (ANN) car-following model to estimate the following vehicle's acceleration based on variable reaction delay input \cite{ModifiedCF2012}. In \cite{ANN2018}, a number of numerical tests showed that ANNs provide a good approximation of car following dynamics. Comparative studies and evaluation between major car-following models under mixed traffic conditions can be found in \cite{Eval2013}.

To describe the driving behavior in various traffic situations, some new methods have been proposed that use mathematical models and neural networks like Bayesian filtering, Recurrent Neural Network to predict a driver’s intended actions across traffic situations \cite{Bayesian2011,RNN2017}. Artificial neural network (ANN) and radial basis function neural network (RBF-NN) showed great benefits to predict the vehicle second-by-second trajectory in congested traffic condition in terms of accuracy and efficiency \cite{PEAD,StopGo2012}. To investigate the cause of stop-and-go pattern and estimate the vehicle behavior in traffic, Agamennoni \textit{et al.} proposed a recursive Bayesian filtering approach for that purpose \cite{Bayesian2011}. The problem of multi-agent inference was tackled by decoupling the joint inference to log-linear combinations of individual dependencies. Toledo \textit{et al.} \cite{TOLEDO2007} further developed an integrated driving behavior model
that combine lane changing behavior and acceleration based on target lane model and target gap model as short-term goal and plan. 

Some cutting-edge research involved studying the interaction between human driven vehicle and CAVs. To investigate the impact of AVs on traffic flow, the authors in \cite{ACC2010} assumed both AVs and human-driven vehicles follow the well-known intelligent driver model (IDM) but with different parameters. Simulation study of a single AV and several human-driven vehicles interaction showed with stabilization can be achieve via a single autonomous vehicle driving around the equilibrium speed \cite{DAN2017}. In this work, a second-order car following model (i.e. optimal-velocity-follow-the leader (OV-FTL) model) was applied to describe the human-driven vehicles' behaviors as well as AV. With estimating vehicle behaviors and anticipate their future trajectories, more effective coordination between vehicles could be achieved in mixed traffic condition.



\section{Coordination of CAVs}\label{CAV}
In this section, we mainly review the literature that were focused on the low-level system of the coordinated ramp control, which is the coordination of CAVs in terms of their motion control algorithms. Some previous work were reviewed by Rios-Torres \textit{et al.} and Scarinci \textit{et al.} \cite{Rios2017survey, scarinci2014control}. However, in this section, we strategically categorize all related literature into two types: centralized approaches and distributed approaches. Additionally, we include more recent papers that were not reviewed by the previous surveys. 




\subsection{Centralized Approaches}
In this section, we define a CAV coordination approach as centralized if the tasks and control commands of the system are globally conducted by the roadside infrastructure and/or the transportation management center (TMC) for all CAVs. Some of the major reasons that such coordination is conducted in a centralized manner are that, each CAV in the system might not have global information of the system, nor can they conduct computations locally that consume large computational power and long computational time. 

In some centralized approaches, tasks and control commands were executed by different layers of the centralized controller. A two-layer CAV coordination system at ramp was proposed by Schimidt \textit{et el.} in 1983, which was based on non-linear system dynamics behavior \cite{schmidt1983atwolayer}. In their system, the higher layer is in charge of the sequence control, while the lower layer is in charge of the motion control of vehicles. Ran \textit{et al.} proposed a similar centralized multi-layer automated ramp system, and also built a microscopic simulation model to validate its characteristics, as well as the designing requirements of the minimum ramp length \cite{ran1999amicroscopic}.

Other than aforementioned approaches, coordination of CAVs at ramp can also be modeled as an optimization problem to be solved by the centralized controller, with the aim of minimizing travel time or fuel consumption. Awal \textit{et al.} proposed an optimization problem with the objective of reducing merging time at ramps and thus reducing merging bottlenecks \cite{awal2013optimal}. Raravi \textit{et al.} formulated an optimization problem which aims at minimizing the Driving-Time-To-Intersection (DTTI) of vehicles, subject to certain constraints for ensuring safety \cite{raravi2007merge}. Rios-Torres \textit{et al.} presented an optimization framework and an analytical closed-form solution that allowed online coordination of CAVs at on-ramp merging zones \cite{riostorres2017automated}. Xie \textit{et al.} formulated this case as a nonlinear optimization problem, and conducted a simulation evaluation with MATLAB and Car2X module in VISSIM \cite{xie2017collaborative}.

\subsection{Distributed Approaches}
Different from centralized approaches that rely on the roadside infrastructure and/or the TMC with infrastructure-to-vehicle (I2V) communication, distributed approaches of CAV coordination at ramp control conduct coordination decisions locally among different CAVs through vehicle-to-vehicle (V2V) communication. Compared to the centralized approaches, the distributed approaches bring several benefits, including reducing communication requirement and improving scalability.

The concept of virtual vehicle coordination at ramps was originated from Uno \textit{et al.} \cite{uno1999amerging}. The proposed approach maps a virtual vehicle onto the highway main line before the actual merging happens, allowing vehicles to perform safer and smoother merging maneuver. Lu \textit{et al.} applied a similar idea in their proposed systems, where they first formulated the merging problems differently with respect to two different geometric layouts of the road (\textit{i.e.}, either with or without a parallel lane), and then proposed a speed based closed-loop adaptive control method to control the longitudinal speed of merging CAVs \cite{lu2000longitudinal}.

Besides the virtual vehicle ideas, other distributed approaches were also proposed to control the longitudinal motion of CAVs at ramps. Dao \textit{et al.} proposed a distributed control protocol to assign vehicles into vehicle strings in the merging scenario \cite{dao2008distributed}. Zhou \textit{et al.} developed a vehicle trajectory planning method for CAV coordination at ramp, formulating the planning tasks of the ramp vehicle and the mainline vehicle as two related distributed optimal problems \cite{zhou2018optimal}. Wang \textit{et al.} proposed a distributed consensus-based CAV coordination system \cite{wang2018distributed}. Furthermore, agent-based modeling and simulation of the proposed CAV coordination system was conducted in game engine Unity3D, and it was compared to human-in-the-loop simulation to evaluate its benefits in terms of mobility and sustainability \cite{wang2018agent, Wang2019Cooperative}.

\section{Conclusions and Discussion}\label{Conclusion}

This paper proposed a system architecture for cooperative ramp control in a mixed traffic environment, and presented a survey on the following key aspects:




\begin{itemize}
    \item \textit{Traffic State Estimation.} We demonstrated a few representative methodologies to estimate traffic states from infrastructure-based surveillance as well as CAVs for the reference of corridor-wide ramp metering. The model-based traffic estimation approaches have been the most commonly used for decades. However, with the improvement in sensing techniques and high-performance computing, plus the advent of CAVs, the research focus has gradually turned to learning-based and/or streaming-data-driven approaches. Moreover, the streaming-data-driven approaches seem to be quite promising, due to their robustness to the disturbances such as non-recurrent traffic situations.
    \item \textit{Corridor-level Control.} From the perimeter control perspective, we illustrated how the most popular ramp metering algorithms, especially those focusing on system-wide benefits, determine the ramp inflow rate in a coordinated manner. The effectiveness of reviewed algorithms has been proven via microscopic simulation and/or real world implementation. Recently, considerable attention has been shifted to learning-based approaches (e.g., Reinforcement Learning), because they showed great potentials to achieving satisfactory solutions without compromising the real-time performance.
    \item \textit{Ramp-level Control.} Keeping in mind the context of mixed traffic for ramp-level control, we first reviewed the driving behaviors of conventional vehicles in response to other vehicles (either non-CAVs or CAVs). Then, we highlighted the studies on cooperative merging protocols for CAVs in scenarios with both full and partial penetration rates. The proper trajectory planning for CAVs in mixed traffic environment is more challenging due to resultant disturbance from human-driven vehicles. Thus, a considerable amount of effort has been made to understand human behaviors when interacting with other traffic. Both microscopic behavior models (\textit{e.g.}, car-following) and machine learning techniques (\textit{e.g.}, ANN) have been well adopted to predict human-driven vehicles' trajectories and/or infer driving intention. Therefore, cooperative ramp control in mixed traffic could be achieved by integrating the predicted conventional vehicles' dynamics into the constraints of the solution space for coordinated trajectory planning of CAVs.
\end{itemize}

Although many positive findings related to coordinated ramp control in mixed traffic scenarios have been reviewed and analyzed in this survey, there are still some open questions that need to be addressed as future work:

\begin{enumerate}
    \item \textit{How to build a more reliable architecture of coordinated ramp control systems in a mixed traffic environment?} A ramp merging area is usually featured as a highly dynamic environment, where driving behaviors may bifurcate significantly due to a variety of factors, such as congestion level, weather condition, and even the drivers' prevailing mood. The presence of CAVs further complicates the situations. The ability to quickly identify the operation scenarios at both corridor level and ramp level, and the flexibility to adapt to different scenarios would be attractive features to enable a reliable ramp control system. Also, due to the involvement of CAVs, V2V communications and I2V communications play critical roles. A resilient system should take into account the odds of packet loss and latency, as well as communication vulnerability subject to cyber-attacks (such as jamming, data injection, and vehicle sensor manipulation).
    \item \textit{How to realize a hybrid ramp control in mixed traffic?} According to the literature review, few studies developed ramp merging control for mixed traffic and even fewer discussed the realization of the proposed system. Minimum effort may involve the upgrade of conventional ramp metering infrastructure with wireless communication capabilities to tally the inflow rates of CAVs for the regulation (with signal head display) of conventional vehicles' inflow rates. CAVs' dynamics are governed by the developed control schemes, regardless of signal head display on the ramp meter. Some safety concerns may arise due to such mixed operation on the same facility (or lane). Another option is to build dedicated infrastructure such as special lanes or grade-separated facilities for preferential treatment of CAVs. However, the system efficiency may be degraded if the demand ratio between CAVs and conventional vehicles does not match the supply (roadway capacity) for respective category. Similar considerations may be applied to the intersection management in presence of both CAVs and conventional vehicles.
    \item \textit{How to balance the system performance and computational effort?} To achieve better performance, higher resolution of data and more delicate control would be preferable. Adaptive strategies on the selection of system parameters (e.g., sensing or control frequency) and system structure (i.e., centralized, decentralized, distributed, and hierarchical) may be necessary in response to various CAV penetration rates, while maximizing the utilization of available resources (for computation or storage).
    \item \textit{How to further boost the system capacity?} It is self-evident that the system capacity varies with the penetration rate of CAVs. In the ideal situation where all vehicles are CAVs, the optimal scheduling strategy can be developed to achieve the maximum throughput of the target ramp merging area. From the perspective of entire corridor, the coordination of inflow rates among ramps is needed to achieve system-wide optimum. In addition, other strategies such as variable speed limit (or intelligent speed adaptation) and cooperative lane changing, when deployed in combination, could help improve the capacity of overall freeway system.  
\end{enumerate}

\section*{Acknowledgment}
This study is funded by the National Center for Sustainable Transportation (NCST). The contents of this paper reflect only the views of the authors, who are responsible for the statement presented herein.


\bibliographystyle{IEEEtran}
\bibliography{Ref.bib}
%

\end{document}